\documentclass[a4paper, oneside, twocolumn, notitlepage, 10pt]{extarticle_ecoc}
\usepackage{ecoc}

\begin{document}
\selectlanguage{english}    


\title{Transoceanic Phase and Polarization Fiber Sensing using Real-Time Coherent Transceiver}%


\author{
    Mikael Mazur\textsuperscript{(1)}, Jorge C. Castellanos\textsuperscript{(2)}, 
    Roland Ryf\textsuperscript{(1)},  
    Erik B{\"o}rjeson\textsuperscript{(3)},
    Tracy Chodkiewicz\textsuperscript{(4)},\\  
    Valey Kamalov\textsuperscript{(2)}, 
    Shuang Yin\textsuperscript{(2)}, 
    Nicolas K. Fontaine\textsuperscript{(1)}, 
    Haoshuo Chen\textsuperscript{(1)}, Lauren~Dallachiesa\textsuperscript{(1)}, \\ 
    Steve~Corteselli\textsuperscript{(1)},
    Philip Copping\textsuperscript{(4)}, J{\"u}rgen Gripp\textsuperscript{(4)}, 
    Aurelien Mortelette\textsuperscript{(4)}, 
    Benoit Kowalski\textsuperscript{(4)}, \\ Rodney Dellinger\textsuperscript{(4)},
    David T. Neilson\textsuperscript{(1)} and Per Larsson-Edefors\textsuperscript{(3)}
}

\maketitle                  


\begin{strip}
 \begin{author_descr}
 
   \textsuperscript{(1)} Nokia Bell Labs, 600 Mountain Ave., Murray Hill, NJ 07974, USA
   \textcolor{blue}{\uline{mikael.mazur@nokia-bell-labs.com}}
   \textsuperscript{(2)} Google LLC, Mountain View, CA, 94043 USA.\\
    \textsuperscript{(3)} Department of Computer Science and Engineering, Chalmers University of Technology, Sweden \\
   \textsuperscript{(4)} Nokia, 600 Mountain Ave., Murray Hill, NJ 07974, USA\\
 \end{author_descr}
\vspace{-15mm}
\end{strip}

\setstretch{1.1}
\renewcommand\footnotemark{}
\renewcommand\footnoterule{}
\let\thefootnote\relax\footnotetext{}

\begin{strip}

  \begin{ecoc_abstract}
    We implement a real-time coherent transceiver with fast streaming outputs for environmental sensing. 
    Continuous sensing using phase and equalizer outputs over 12800\,km of a submarine cable is demonstrated to enable time resolved spectroscopy in broad spectral range of 10\,mHz - 1\,kHz. \vspace{-3mm} 
  \end{ecoc_abstract}
\end{strip}


\section{Introduction}
\label{sec:intro}
Sensing over live telecom networks has an enormous potential due to the vast extent of deployed fibers. 
It is highly desirable to add sensing capabilities to coherent transceivers to avoid the cost associated with dedicating wavelength channels or dedicated dark fibers. 
While geophysical sensing of earthquakes and water waves have been demonstrated using real-time transceiver readouts of the state of polarization (SoP) over submarine cables~\cite{Zhan2021}, and phase sensing using offline processing has been demonstrated in metro networks~\cite{Wellbrock2021}, 
real time read out of phase information in transceivers has yet to be demonstrated.

Here we demonstrate 
the first real-time coherent transceiver with built-in real-time phase and polarization sensing while simultaneously transmitting information. We successfully transmit over 12,800\,km while continuously performing environmental sensing. The DSP implementation is designed for transmission and the sensing is enabled via pipelined extraction using RAM memory as fast bridges between the DSP and external CPU clock domain, supporting dynamically re-configurable readout speed up to $~$\,MHz and data decimation.
We first show sensing using standard polarization measurements and verify our results with a commercial transceiver. The events are then analyzed using additional sensing metrics based on the full Jones matrix. This significantly improves the sensitivity and reveals additional event features, not detectable using traditional SoP measurements. 
Our results show the potential of real-time coherent transceivers to perform sensing, enabling environmental monitoring and advanced network management. 

This work is based on the pioneering contribution of G.~Marra et al.~\cite{Marra2018}, where interferometric laser-based phase measurements were used to sense an earthquake.
Real-time coherent transceiver based fiber sensing complements distributed acoustic sensing based on Rayleigh backscattering, Raman or Brillouin scattering, which are
well known tools to transform the fiber grid into highly precise environmental sensors~\cite{Lindsey2019}. 

\section{Real-Time Transceiver Implementation}
\label{sec:tranceiver}
\begin{figure*}[t]
   \centering
    \includegraphics[width=.9\linewidth]{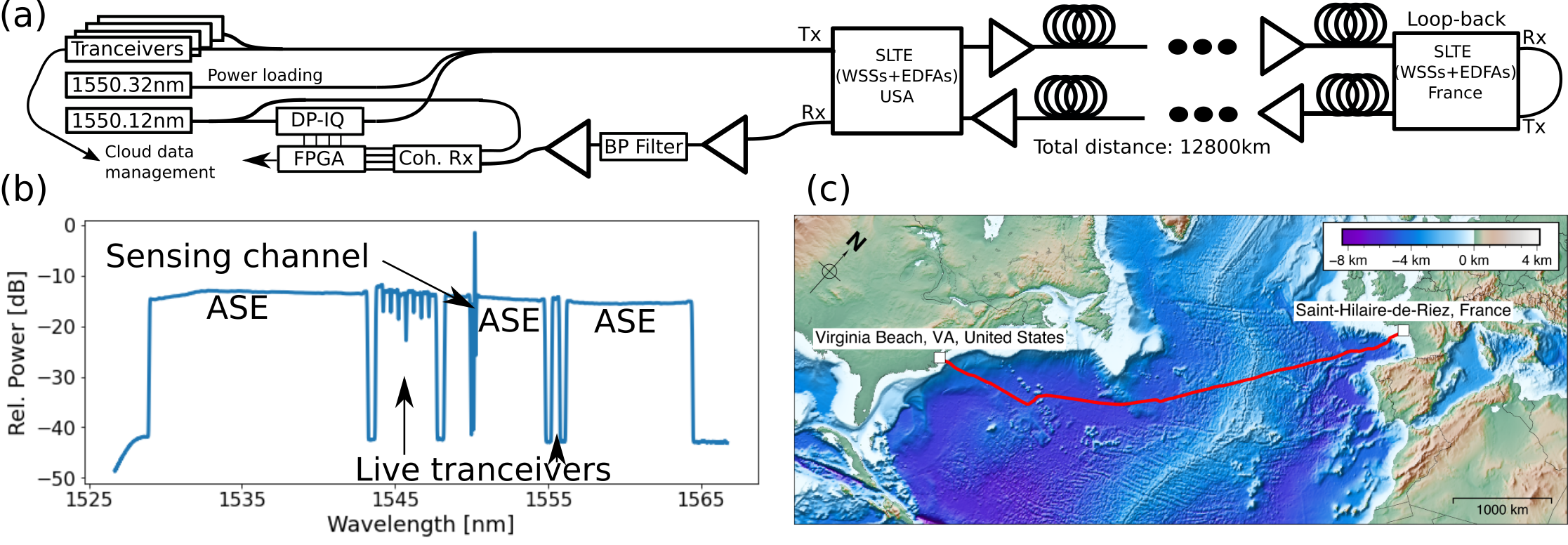}
    \caption{(a) Experimental setup for coupling the sensing test channel to the cable. (b) Transmitted spectrum including test channel, nine live transceivers and ASE. (c) Regional map showing the geometry of the Dunant fiber-optic cable (red line)}
    \label{fig:setup}
\end{figure*}
The FPGA-based transceiver was implemented on a Xilinx RF-SoC (ZU49DR) containing logic resources for real-time processing, analog-to-digital/digital-to-analog converters (ADCs/DACs) and a CPU. 
The ADCs are sampled at 2\,GS/s with 8-bit resolution. A DSP clock of 125\,MHz, resulting in a parallelization degree of 8 symbols (16 input samples), was used for the DSP. 
A $2\times2$ complex-valued multiple-input-multiple-output (MIMO) equalizer with 17 T/2-spaced taps updated by a constant modulus algorithm was used for equalization and polarization de-multiplexing. 
The equalizer used 9-bit resolution for both coefficients and error calculation. 
The carrier phase estimation (CPE) consisted of a principal component based phase tracker\cite{Borjeson2021}, using 7 bits angle resolution and 64 symbols block averaging. 
To enable combined parallel DSP and environmental sensing, information must flow in real time from the hardware level (clocked logic) to the CPU and for post-processing/data management. 
To solve this, we use internal tracking registers built into the DSP blocks, dedicated pipelines and parallel dual-port RAM interfaces to bridge the fast DSP and slow CPU clock domains. We output all complex taps (72 complex values) together with the amount of tracked phase per polarization. 
The output speed is re-configurable with a maximum of $\approx$MHz, enabling dynamic tuning of the interferometer frequency range.

\section{Experimental Setup}
\begin{figure}[t]
   \centering
     \includegraphics[width=1.\linewidth]{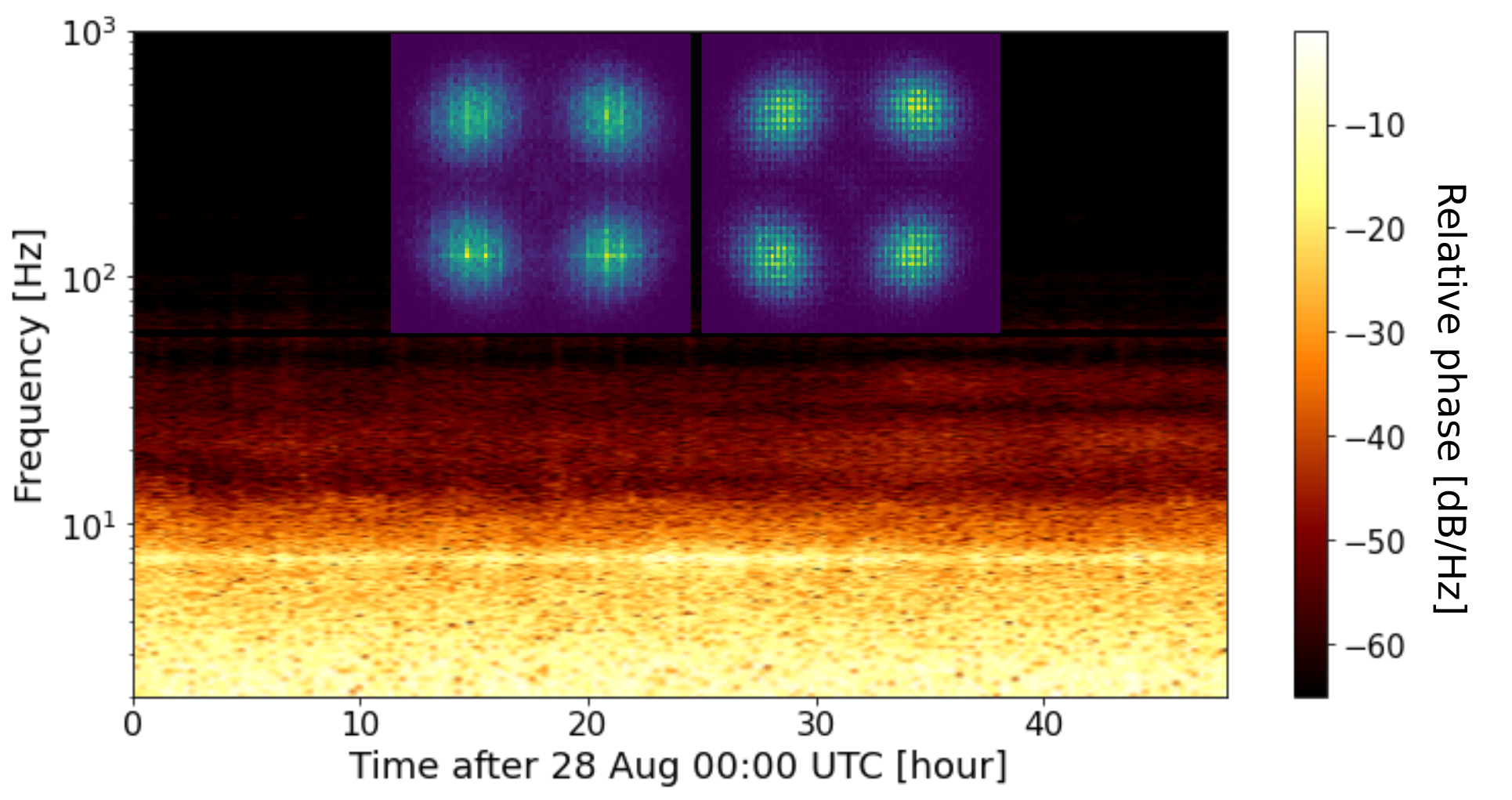}
    \caption{Sensing interferometer spectrogram over 48 hours. The phase is reconstructed by combining phase and polarization data continuously extracted form the real-time DSP. 
    Inset shows constellation diagrams for x-/y-polarization. \vspace{-2mm}} 
    \label{fig:spectrum}
\end{figure}
A schematic of the experimental setup is shown in Fig.~\ref{fig:setup}(a). 
A dual-polarization Mach-Zehnder IQ-modulator driven by four DACs from the FPGA was used to modulate a 1\,GBd QPSK generated using decorrelated PRBS15 signal on a low linewidth laser. Due to the narrow band signal, a CW laser was placed about 25\,GHz away from the test channel to give the slot equivalent power. 
The transmitted spectrum is shown in Fig.~\ref{fig:setup}(b).
The signal was injected into the submarine line terminal equipment (SLTE), containing a twin 1x32 WSS, and combined with additional commercial channels and amplified spontaneous emission (ASE) noise to fill the total bandwidth supported by the cable.  The signal was then injected into 1 fiber pair of the Dunant cable, shown in Fig.~\ref{fig:setup}(c) connecting Virginia Beach in the United States to Saint-Hilaire-de-Riez in France. The cable supports 12 fiber pairs and the total distance is 6,400\,km. In France, an identical SLTE was used to extract the signal and send it to the input of the other fiber in the pair. 
The received signal was selected using the SLTE and a two stage amplifier with a 0.2\,nm wide bandpass. 
The output was combined with a tap from the transmitter laser (''homodyne'' configuration) on a coherent receiver and processed by the FPGA. An example 48 hours spectrogram, created by extracting phase information from the equalizer and phase tracker together with the x-/y constellations, both after 12,800\,km, is shown in Fig.~\ref{fig:spectrum}. The signal-to-noise ratio was estimated to 10\,dB.   

 \vspace{-2mm}
\section{Polarization and Phase Sensing}
The SoP and phase information resides in both the equalizer and CPE blocks. While the phase tracking information is directly extracted using registers counting the amount of applied phase, extracting the SoP and phase from the equalizer is more challenging. 
The $2\times2$ MIMO equalizer filter separating x and y can be expressed as\cite{Zhan2021,Savory2010}
\vspace{-4mm}
\begin{equation}
\vspace{-2mm}
\label{eq:jones}
    H_{\text{inv}}(\omega) =  
    \begin{bmatrix}
    H_{\text{xx}}(\omega) &  H_{\text{xy}}(\omega) \\ 
    H_{\text{yx}}(\omega) &  H_{\text{yy}}(\omega)
    \end{bmatrix}.
\end{equation}
Previous work on geophysical sensing using coherent transceivers has mainly focused on using a Stokes vector ($S_1$,$S_2$,$S_3$). A Stokes vector can be derived using either of the two rows of Eq.~\ref{eq:jones}~\cite{Zhan2021}.
This implies that a perturbation causing an absolute phase difference between these two vectors would not be captured using a single Stokes vector. Frequency dependence is also neglected. 
A Jones matrix therefore contains more ''sensing'' information than a single Stokes vector\cite{Mecozzi2021}. 
To remove the $\leq$2\,dB PDL from the measured Jones matrix, Eq.~\ref{eq:jones}, we follow the process outlined in\cite{Karlsson2000} and decompose it into one hermitian and one unitary matrix, describing the PDL and SoP rotation, respectively.
After extracting the unitary (rotation) part, the correlation, or closeness, of two unitary Jones matrices can be calculated as $C = \frac{\text{Tr}\left(U_2^HU_1\right)}{2}$, with $U_2,U_1$ denoting the two Jones matrices~\cite{Mazur2018d}. Aside from being phase sensitive, it is independent of the SoP, removing any effects from static rotations~\cite{Mazur2018d}. 

\section{Results and Conclusion}
\begin{figure}[t]
   \centering
    \includegraphics[width=1.\linewidth]{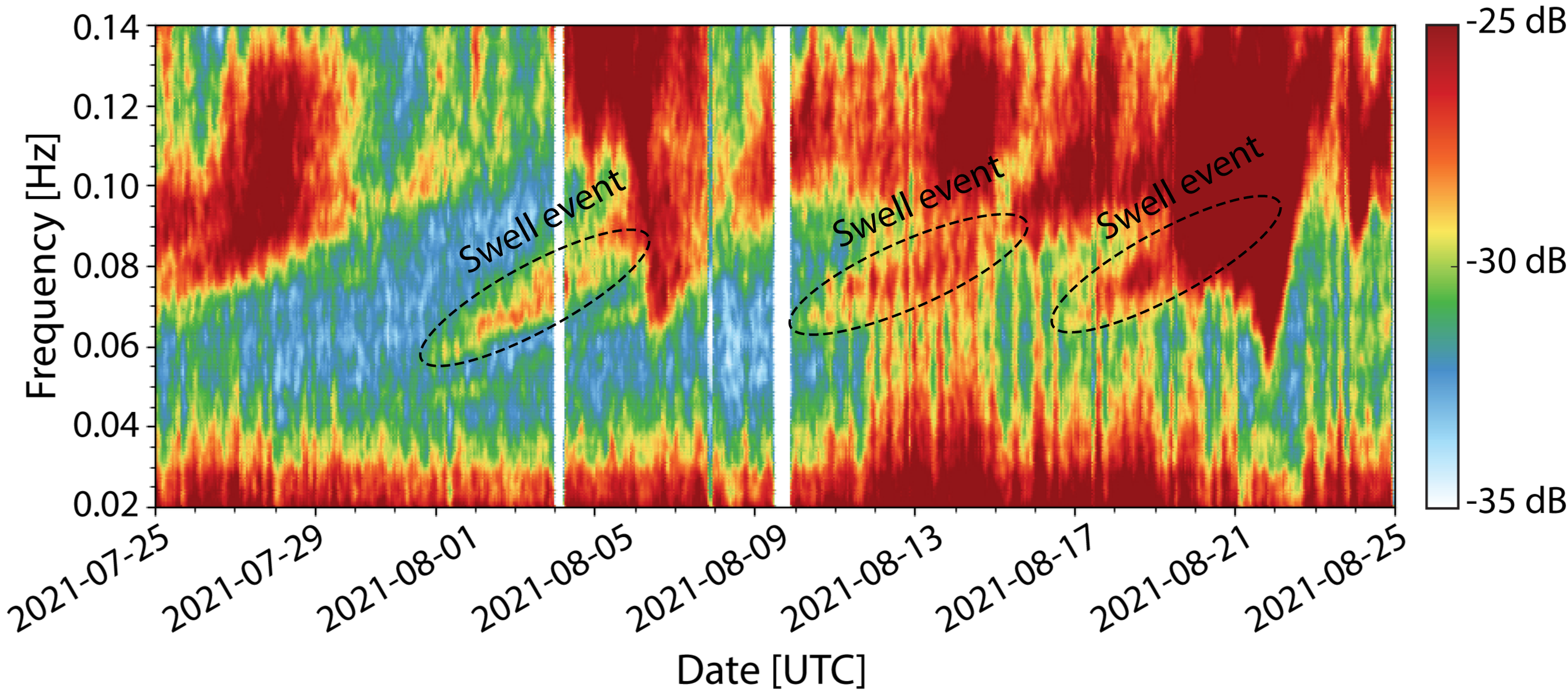}
    \caption{Spectrogram of average SoP rotation between $S_1$ and $S_2$ after rotation to $S_3=1$ over 1 month of measurements on the Dunant cable. The dispersive energy packets around 0.06-0.10 Hz, in the primary microseism band, are caused by the water wave pressure that is exerted by distant storms.\vspace{-2mm}}
    \label{fig:res1}
\end{figure}
\begin{figure}[t]
   \centering
    \includegraphics[width=1.\linewidth]{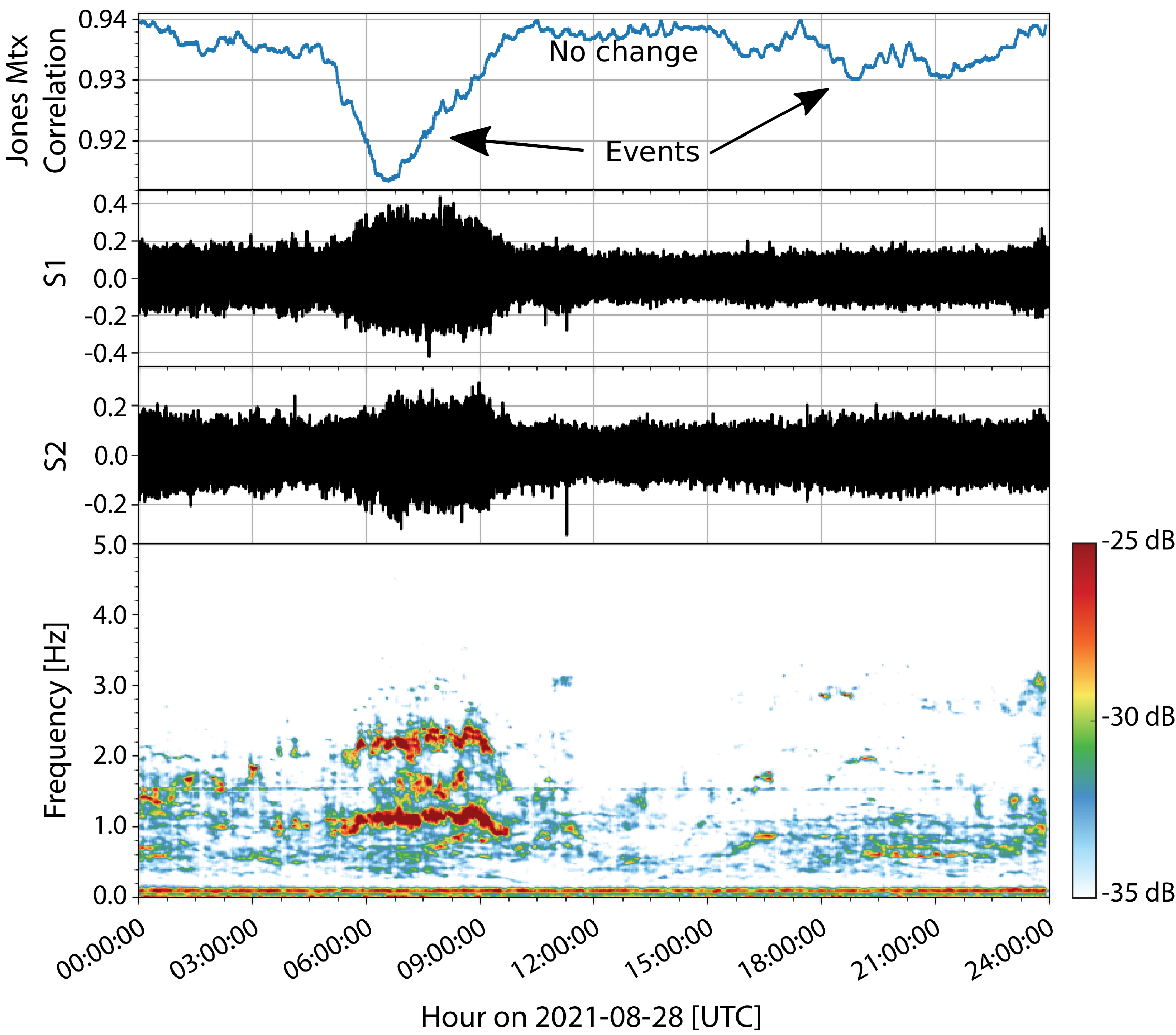}
    \caption{Correlation $C$ between measured Jones matrices accounting for both SoP rotation and relative phase change (top),  $S_1$ and $S_2$ changes after projection to $S_3=1$ (middle) and associated spectrogram (below).  \vspace{-3mm}}
    \label{fig:res2}
\end{figure}

 
As a first transceiver-based sensing example, we analyze frequencies up to 0.14\,Hz, covering the primary and secondary microsesmic bands\cite{Ardhuin2019}. 
Figure~\ref{fig:res1} shows a one-month spectrum of the Dunant SoP fluctuations. It reveals several bursts of energy lasting a few days located in the primary band, around 0.06\,Hz.
Strong winds associated with tropical cyclones exert intense basal stresses on the ocean surface and generate large amplitude water waves (swells) capable of propagating large distances with minimal attenuation. 
Because low-frequency swells propagate faster than high-frequency, their frequency dependence, or temporal tilt in Fig.~\ref{fig:res1}, can be used to infer the waves' origin\cite{Gualtieri2020}. 
These events correspond to pressure signals arising from distant pelagic storms and passing by the submarine cable, as confirmed by their notable dispersive behavior that is characteristic of water waves traveling in deep water\cite{Ardhuin2015}.
Direct observation of swells is normally poor due to the sparse sensor instrumentation across the world's ocean. Converting transoceanic telecom cables to environmental sensors will, therefore, permit rapid detection and location of storms throughout the globe. 
Since the amplitude of these waves is comparable to those of tsunamis in the open ocean, it shows the potential for sensing transceivers to be used for tsunami early warning systems. 
To further enhance the sensitivity, we add the relative phase between the two orthogonally modulated polarizations, which creates an additional interferometer that is highly sensitive to fiber perturbations. 
Figure~\ref{fig:res2} shows the measured correlation $C$, the SoP (both $S_1$ and $S_2$) together with a spectrogram and over 24 hours. 
We observe an increased rotation amplitude between 6:00 and 9:00am. It appears to be related to a resonance of the system, as evidenced by their double-frequency peaks at around 1 and 2\,Hz. 
The Jones-matrix correlation, $C$, reveals additional information about the event, such as the fact hat the rotation speed increased more rapidly than it decreases. 
While not conclusive, we believe that this event could be related to either acoustic waves trapped in the ocean layer, or to some cable strumming phenomenon excited by deep water currents. 
For events like these, and many other unidentified events, we believe that the additional information provided by complementing SoP rotations with phase is vital for accurate identification and categorization. 
Our prototype also resolves features beyond the sensitivity of regular SoP measurements occurring between 15:00 and midnight. 
Directly comparing SoP measurements from a commercial transceiver and the matrix correlation in Fig.~\ref{fig:res2}, we find a remarkable agreement with both timing and amplitude.
We interpret this as a strong indicator that the sensing capabilities of our prototype are adequate to detect mechanical perturbations along the cable and is, therefore, a complementary improvement on the use of telecom networks for monitoring seismic activity and ocean dynamics.


In conclusion, coherent transceivers revolutionized fiber optics communications, thanks to the ability to unscramble the polarization states and track phase of the transmitted signal using digital signal processing. 
In this work, we show that tracking the complex equalizers coefficients and monitoring the phase in real time can transform a conventional transceiver into a powerful sensing device, capable of sensing the full Jones matrix of the transmitted channel and offering enhanced sensing capabilities over conventional SoP measurements, with wide-ranging applications including terrestrial networks and submarine links.






\newpage
\printbibliography
\vspace{-4mm}

\end{document}